\documentclass[aps,reprint,superscriptaddress]{revtex4-2}
\usepackage{siunitx}
\usepackage{graphicx}
\usepackage{placeins}
\usepackage{textcomp}
\usepackage{amssymb}
\usepackage{makecell}
\usepackage{physics}

\graphicspath{{Figures/}}
\begin{document}

\title{Fast optical control of a coherent hole spin in a microcavity}

\author{Mark R. Hogg}
\thanks{These two authors contributed equally}
\author{Nadia O. Antoniadis}
\thanks{These two authors contributed equally}
\affiliation{Department of Physics, University of Basel, Klingelbergstrasse 82, CH-4056 Basel, Switzerland}
\author{Malwina A. Marczak}
\affiliation{Department of Physics, University of Basel, Klingelbergstrasse 82, CH-4056 Basel, Switzerland}
\author{Giang N. Nguyen}
\affiliation{Department of Physics, University of Basel, Klingelbergstrasse 82, CH-4056 Basel, Switzerland}
\author{Timon Baltisberger}
\affiliation{Department of Physics, University of Basel, Klingelbergstrasse 82, CH-4056 Basel, Switzerland}
\author{Alisa Javadi}
\altaffiliation{Present address: School of Electrical and Computer Engineering, Department of Physics and Astronomy, The University of Oklahoma, 110 West Boyd Street, OK 73019, USA}
\affiliation{Department of Physics, University of Basel, Klingelbergstrasse 82, CH-4056 Basel, Switzerland}
\author{R\"{u}diger Schott}
\affiliation{Lehrstuhl f\"{u}r Angewandte Festk\"{o}rperphysik, Ruhr-Universit\"{a}t Bochum, D-44780 Bochum, Germany}
\author{Sascha R. Valentin}
\affiliation{Lehrstuhl f\"{u}r Angewandte Festk\"{o}rperphysik, Ruhr-Universit\"{a}t Bochum, D-44780 Bochum, Germany}
\author{Andreas D. Wieck}
\affiliation{Lehrstuhl f\"{u}r Angewandte Festk\"{o}rperphysik, Ruhr-Universit\"{a}t Bochum, D-44780 Bochum, Germany}
\author{Arne Ludwig}
\affiliation{Lehrstuhl f\"{u}r Angewandte Festk\"{o}rperphysik, Ruhr-Universit\"{a}t Bochum, D-44780 Bochum, Germany}
\author{Richard J. Warburton}
%\email[To whom correspondence should be addressed:]{richard.warburton@unibas.ch}
\affiliation{Department of Physics, University of Basel, Klingelbergstrasse 82, CH-4056 Basel, Switzerland}

\date{\today}

\begin{abstract}
A spin-photon interface is one of the key components of a quantum network. Physical platforms under investigation span the range of modern experimental physics, from ultra-cold atoms and ions to a variety of solid-state systems. Each system has its strengths and weaknesses, typically with a trade-off between spin properties and photonic properties.  
Currently, the best deterministic single-photon sources use a semiconductor quantum dot embedded in an optical microcavity. However, coherent spin control has not yet been integrated with a state-of-the-art single-photon source, and the magnetic noise from host nuclear spins in the semiconductor environment has placed strong limitations on the spin coherence.
Here, we combine high-fidelity all-optical spin control with a quantum dot in an open microcavity, currently the most efficient single-photon source platform available.
By imprinting a microwave signal onto a red-detuned optical field, a Raman process, we demonstrate coherent rotations of a hole spin around an arbitrary axis of the Bloch sphere, achieving a maximum $\pi$-pulse fidelity of 98.6\%. The cavity enhances the Raman process, enabling ultra-fast Rabi frequencies above 1~GHz. We use our flexible spin control to perform optical cooling of the nuclear spins in the host material via the central hole spin, extending the hole-spin coherence time $T_{2}^{*}$ from 28\,ns to 535\,ns. Hahn echo preserves the spin coherence on a timescale of 20~$\mu$s, and dynamical decoupling extends the coherence close to the relaxation limit. 
Both the spin $T_{2}^{*}$ and spin rotation time are much larger than the Purcell-enhanced radiative recombination time, 50~ps, enabling many spin-photon pairs to be created before the spin loses its coherence.
\end{abstract}

\maketitle

\begin{figure*}[t]
    \centering
    \includegraphics[width = \linewidth]{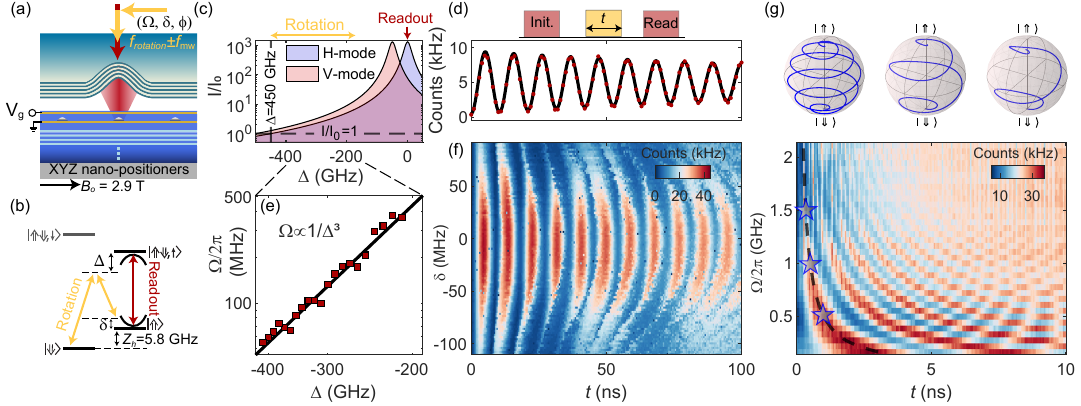}
        \caption{\textbf{All-optical coherent control of a quantum dot hole spin in a microcavity.} 
    \textbf{(a)} Illustration of the spin control scheme. A laser resonant with the microcavity (red) is used for spin readout and initialisation, and a red-detuned laser (yellow) for full spin control via a Raman process (with Rabi frequency $\Omega/(2\pi)$, detuning $\delta$ and phase $\phi$).
    \textbf{(b)} QD energy levels and transitions. The spin states of a single hole define our qubit; the strong optical trion transitions are used for spin initialisation, readout and control. The readout laser is resonant with the the $\ket{\Uparrow}\leftrightarrow\ket{\Uparrow\Downarrow,\uparrow}$ transition; the Raman control laser is detuned from resonance by $\Delta$.
    \textbf{(c)} Optical intensity enhancement as a function of $\Delta$ for our microcavity. The cavity provides an enhancement until $\Delta=450$\,GHz, perfectly suitable for spin control. The cavity linewidth is $\kappa/(2 \pi)=25$~GHz, the mode splitting 50~GHz.
    \textbf{(d)} Hole spin Rabi oscillations: readout signal as a function of rotation laser pulse length $t$ for $\Delta=320$\,GHz, with $\Omega/(2\pi)=95$~MHz.
    \textbf{(e)} Rabi frequency $\Omega/(2 \pi)$ as a function of the Raman laser detuning $\Delta$ at constant laser power. $\Omega$ has a $1/\Delta^{2.99\pm0.14}$ dependence. 
    \textbf{(f)} Rabi oscillations as a function of control laser Raman detuning $\delta$, exhibiting a characteristic chevron pattern. 
    \textbf{(g)} Ultra-fast Rabi oscillations for $\Delta=150$\,GHz, achieving Rabi couplings of $\Omega/(2\pi)>2$\,GHz. Bloch spheres: simulated Bloch vector trajectory in the lab frame for a $\pi$-pulse (indicated by blue stars) with $\Omega/(2\pi)=500$\,MHz (left), 1\,GHz (center) and 1.5\,GHz (right).
    }
    \label{concepts}
\end{figure*}

Generating entanglement between flying photonic qubits and stationary matter qubits (such as a coherent spin) is a key challenge in quantum information science \cite{DiVincenzo2000, Kimble2008}. Spin-photon entanglement provides the basis for a distributed quantum network, and entangled multi-photon cluster states can be generated using the stationary qubit as an entangling mediator \cite{Lindner2009, Thomas2022, Tiurev2022}. More complex graph states can be created, either by fusion or by gates between matter qubits, and represent the central resource for photonic measurement-based quantum computing \cite{Raussendorf2001,Raussendorf2003,Briegel2009}, as well as all-optical quantum repeaters \cite{Azuma2015,Buterakos2017,Borregaard2020}. Promising platforms include single atoms in vacuum \cite{Wilk2007, Ritter2012, Thomas2022}, colour centres in diamond \cite{Togan2010, Bernien2013, Stas2022, Knaut2024}, and semiconductor quantum dots (QDs) \cite{Gao2012, Schwartz2016, Appel2022, Coste2023, Cogan2023, Meng2023, Meng2023_fusion}. As minimum requirements, the spin should retain its coherence as the photons are created, and the photons should be extracted with high efficiency. Additionally, a high rate of entanglement generation is desirable in order to run protocols at high clock-rates, while also minimising the path length and storage time requirements for quantum optical technologies. 

In terms of deterministic single-photon sources, semiconductor QDs stand out. The coherence (as judged by the two-photon interference visibility) of successively emitted photons is high, is maintained over long photon streams, and even on employing photons created by separate QDs \cite{Zhai2022}. The extraction efficiency can be high by engineering the photonic environment with various means \cite{Santori2002,Somaschi2016,Wang2019PRL,Liu2018,Arcari2014,Uppu2020,Liu2019,Barbour2011,Tomm2021} such that the majority of photons are emitted into one mode. The system losses can be low: end-to-end efficiencies exceeding 50\% has been demonstrated \cite{Tomm2021, Ding2023}. Moreover, if a microcavity is used for the photonic engineering, the Purcell effect reduces the radiative lifetime to values of 50~ps or below \cite{Liu2018,Tomm2021} such that a QD-based single-photon source can operate at GHz repetition rates. For many years, the Achilles heel of the QD system was the poor electron-spin coherence: magnetic noise from the host nuclei over a wide range of frequencies limits both the coherence time -- $T_{2}^{*}$ is just a few nanoseconds -- and the power of dynamical decoupling \cite{Stockill2016}. Switching to a hole spin increases the $T_{2}^{*}$ by one or possibly two orders of magnitude \cite{Brunner2009,DeGreve2011,Prechtel2016,Huthmacher2018} (provided the electrical noise is low \cite{Prechtel2015}) as a hole spin interacts both weakly and anisotropically \cite{Fischer2008,Prechtel2016} with the host nuclear-spins. More recently, strategies to improve QD electron-spin coherence by narrowing (i.e., cooling) the nuclear spin distribution have been developed \cite{Ethier-Majcher2017, Gangloff2019, Jackson2022}, resulting in electron spin $T_{2}^{*}$ times of up to 600~ns \cite{Nguyen2023}. Also, dynamical decoupling was highly successful on low-strain GaAs QDs for which the magnetic noise lies at relatively low frequencies \cite{Zaporski2023}. Furthermore, the spin can be rotated around the Bloch sphere in just a few nanoseconds by exploiting the strong optical transition, i.e., via a Raman process \cite{Press2010,Bodey2019,Nguyen2023}. The net result is that an electron spin in both strained QDs (e.g.\ InGaAs QDs in GaAs) and low-strain QDs (e.g.\ GaAs in AlGaAs) can retain its coherence as many photons are created, meeting the prerequisite for the creation of high-fidelity cluster states.

So far, QD microcavities and QD spin control have been developed separately. In particular, nuclear spin cooling has been developed on bulk-like structures with low end-to-end single-photon efficiencies \cite{Gangloff2019,Jackson2021,Nguyen2023}. The key challenge is to combine the two. Here, we achieve the best of both worlds: we implement high-fidelity coherent control of a QD spin; the QD is in an engineered photonic environment with a very high end-to-end single-photon efficiency. The spin is a hole spin; the photonic engineering relies on an open microcavity \cite{Barbour2011, Tomm2021}. The hole spin coherence is extended from $T_{2}^{*}=28$~ns to 535~ns by using the hole as a central spin to cool the nuclear spins. Spin control uses a Raman process with two laser fields detuned from the microcavity. We show that this approach is directly compatible with the resonant microcavity, allowing us to integrate spin control with a system engineered for optimal single-photon source efficiency. In fact, we find that the cavity enhances the Raman fields such that Rabi couplings above 2.0~GHz ($\sim34\%$ of the Larmor frequency) are achieved.

\section{Spin control in a microcavity}
Our system consists of an electrically gated InGaAs QD coupled to an open microcavity \cite{Tomm2021, Najer2019} (see Methods and Supplementary Note 1). 
% The QD is in the electron Coulomb-blockade regime. 
A single metastable hole is created on resonant optical excitation by applying an appropriate bias to the device. The planar bottom mirror is part of the semiconductor heterostructure; the curved top mirror is patterned on a silica substrate using laser ablation. A QD can be tuned into resonance with the microcavity in situ using nano-positioners. The same microcavity-coupled QD device was used previously to create a single-photon source with an end-to-end efficiency of 57\% and GHz repetition rates \cite{Tomm2021}. 

A magnetic field of 2.9~T is applied perpendicular to the growth direction of the QDs (Voigt geometry). Figure~\ref{concepts}(a),(b) shows a schematic of the cavity setup and the level structure of the QD. The magnetic field splits the ground state into two, $\ket{\Uparrow}$,$\ket{\Downarrow}$, with Zeeman splitting $Z_h= 5.8$~GHz (hole g-factor $g_h=0.143$). 
% The microcavity is tuned into resonance with the  ``vertical" $\ket{\Uparrow}$-transition. 
A laser resonant with the $\ket{\Uparrow}\leftrightarrow\ket{\Uparrow\Downarrow,\uparrow}$ trion transition (red arrow) is used to initialise the spin to $\ket{\Downarrow}$ via optical pumping \cite{Atature2006,Xu2007}, and to probe the $\ket{\Uparrow}$ population for spin readout (Supplementary Note 3). Spin control proceeds by a two-photon Raman pulse, red-detuned from the initialisation/readout  laser. The two frequencies in the Raman pulse are generated by microwave modulation (with frequency $f_{mw}$ and phase $\phi_{mw}$) of a CW laser, a scheme  providing multi-axis control of the spin state on the Bloch sphere \cite{Bodey2019}.

The microcavity has a linewidth of $\kappa/(2 \pi)= 25$~GHz and a finesse $F=500$. To create photons in the microcavity via the Purcell effect, the microcavity must be resonant with one of the QD transitions. However, coherent spin control requires Raman laser detunings satisfying the criterion $\Delta \gg \Omega_R/(2 \pi)$ to avoid populating the trion states, which randomises the spin when the trion decays. Here, $\Delta$ is the detuning of the Raman laser and $\hbar \Omega_R$ is the optical Rabi coupling of the two Raman fields, assumed to be equal. The spin Rabi frequency is then given by $\Omega/2\pi=(\Omega_R/2\pi)^2/\Delta$. In practice, detunings of $\Delta \sim100-1,000$~GHz are typically used \cite{Press2010, Bodey2019, Nguyen2023}. A potential challenge for spin control in a cavity is thus how to couple the detuned Raman fields to the QD as the QD resides in a narrowband device, the microcavity.

Figure~\ref{concepts}\,(c) shows the optical intensity enhancement provided by the microcavity as a function of $\Delta$. On resonance, the cavity enhances the input optical intensity by a factor of $8F/\pi=1270$ (Supplementary Note 2); the intensity enhancement decreases with detuning and reaches one at a detuning of $\Delta=\frac{\kappa}{2\pi}\sqrt{2F/\pi}\simeq450$~GHz (Fig.~\ref{concepts}(c)). Hence, for $\Delta \lesssim 450$~GHz, the microcavity actually enhances the Raman fields. In other words, the Raman fields couple effectively to the microcavity via the Lorentzian ``tails" of the microcavity-mode. 

We drive coherent rotations between $\ket{\Uparrow}$ and $\ket{\Downarrow}$ with Raman pulses of increasing pulse length $t$, a Rabi experiment, initially with $\Delta=320$~GHz, Fig.~\ref{concepts}(d). Rabi oscillations are observed with quality factors of up to 35 (at a hole spin  Rabi frequency of $\Omega/2\pi=95$\,MHz), corresponding to an upper-bound rotation fidelity of 98.6\%. 
For a fixed laser power, we then measure $\Omega/2\pi$ as a function of $\Delta$, Fig.~\ref{concepts}(e). Fitting the data we find that $\Omega$ scales as $1/\Delta^{2.99\pm0.14}$. In a photonic-broadband device, $\Omega$ scales as $1/\Delta$ such that the extra factor of $1/\Delta^2$ can be attributed to the microcavity, exactly the scaling expected for a Lorentzian spectral-function. On detuning the microwave frequency of the Rabi drive from the Zeeman frequency ($\delta=2f_{mw}-Z_h$), we observe the typical chevron pattern, Fig.~\ref{concepts}(f). As we will show, departures from the expected chevron for an ideal two-level system are due to an interaction between the hole spin and the nuclei of the host material.

The cavity enhancement of the Raman fields allows ultra-fast coherent spin control at modest laser power. To demonstrate this, Fig.~\ref{concepts}(g) shows Rabi oscillations as a function of the Raman laser power for $\Delta =150$~GHz. We observe coherent oscillations with Rabi frequencies greater than $\Omega/(2\pi)=1$~GHz, Fig.~\ref{concepts}(g), i.e., a $\pi$ rotation is achieved in 0.5~ns. At even higher Rabi couplings, additional fast oscillations appear in the Rabi experiment, evidence that the rotating-wave approximation starts to break down.

\vspace{-0.5cm}
\section{Hole spin coherence}
\begin{figure}[hb!]
    \centering
    \def\svgwidth{\textwidth}
    \includegraphics[width = \textwidth]{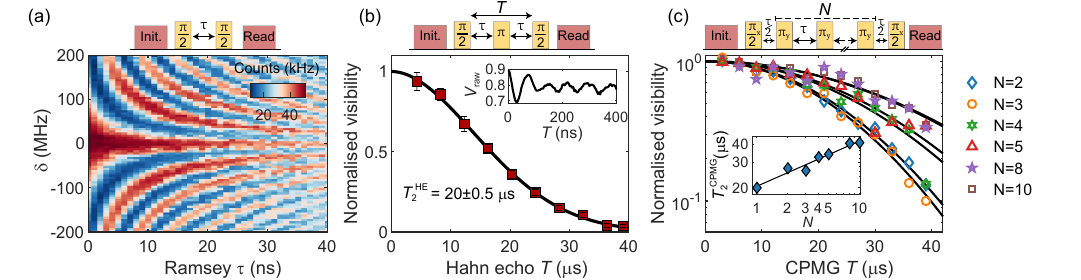}
    \caption{\textbf{Hole spin coherence with thermal nuclear ensemble.}
    \textbf{(a)} Ramsey experiment: readout signal as a function of pulse spacing $\tau$ between two $\frac{\pi}{2}$-pulses ($x$-axis) and $\delta$, the detuning from $Z_h$ ($y$-axis). A $T_{2}^{*}$ of $28\pm2$\,ns is extracted from a Gaussian fit at $\delta=0$. This data was taken at 2.5\,T, in contrast to all other datasets where 2.9\,T was used. Measurements at 2.9\,T give the same $T_2^*$ within the error bounds.
    \textbf{(b)} Hahn echo visibility versus precession time $T$ with a Gaussian fit. We find $T_2^\text{Hahn}=20\pm0.5\,\mu$s. Inset: Hahn echo modulation for short delays, showing a maximum (un-normalised) echo revival visibility of $\sim87\%$, indicating efficient decoupling from noise due to nuclear Larmor precession.
    \textbf{(c)} CPMG dynamical decoupling for different pulse number, $N$. CPMG is effective in extending the hole spin phase coherence from $20\,\mu s$ up to $40\,\mu s$. Inset: $T_{2}^{\text{CPMG}}$ versus CPMG pulse number; the fit is $T_{2}^{\text{CPMG}}=T_{2}^{\text{HE}} \cdot N^{\gamma}$ with $\gamma=0.31\pm0.04$.
    }
    \label{fig:coherence}
\end{figure}

Next, we benchmark the coherence properties of our hole spin. We first measure the $T_{2}^{*}$ time with a Ramsey sequence, extracting the $\frac{\pi}{2}$-pulse length from the Rabi oscillations. Figure~\ref{fig:coherence}(a) shows the readout signal for the Ramsey sequence as a function of $\delta$, where as expected we observe oscillations in the Ramsey signal exactly at frequency $\delta$. From the data at $\delta=0$ we extract a coherence time of $T_{2}^{*}=28\pm2$~ns. As we show below, the mechanism limiting $T_{2}^{*}$ is nuclear spin noise. To avoid polarising the nuclei in the Ramsey experiment, we interleave two Ramsey sequences, one with a phase of $\phi=0$ applied to the second $\frac{\pi}{2}$ pulse, the other with a phase of $\phi=\pi$, such that the average hole spin $z$-projection throughout the pulse sequence is zero \cite{Stockill2016}. We note that applying rotation pulses with arbitrary phase is trivial with the microwave-modulated spin control scheme.

\onecolumngrid

\vspace{\columnsep}
\vspace{\columnsep}
\twocolumngrid

We then perform a Hahn echo to filter the effects of low-frequency noise on the hole spin, Fig.~\ref{fig:coherence}\,(b). The Hahn echo is highly effective, extending the coherence time to $T_2^\text{HE}=20\pm0.5\,\mu$s. We note that $T_2^\text{HE}$ is comparable to the $T_1$ relaxation time (Supplementary Note 4). To isolate the effect of loss of phase coherence from spin relaxation we keep a constant spacing between the initialisation and readout pulses as we scan $T$ in the Hahn echo. In this way spin relaxation has a constant effect on the echo visibility, and loss of visibility as a function of $T$ is due to loss of phase coherence. In other words, we are probing the hole spin pure dephasing rate. We also perform a CPMG sequence, which further extends the phase coherence time (Fig.~\ref{fig:coherence}~(c)). We find that $T_{2}^{\text{CPMG}} = T_2^\text{HE}\cdot N^{\gamma}$, where $N$ is the number of $\pi$-pulses in the CPMG sequence and the scaling exponent $\gamma=0.31\pm0.04$. Previous results for an InGaAs hole spin measured $\gamma=0.325\pm0.005$, limited by charge noise with a $1/f^{0.48}$ frequency spectrum \cite{Huthmacher2018}. The previously reported value of $\gamma$ is remarkably consistent with our current work, however we achieve an order of magnitude longer $T_{2}^{\text{CPMG}}$ (40\,$\mu$s compared to 4\,$\mu$s), likely due to a lower level of charge noise. We achieve this by rigorously controlling the purity of our MBE grown material \cite{Nguyen2020} and by designing our heterostructure without the presence of artificial charge trap states \cite{Ludwig2017}.

\section{Nuclear Spin cooling via a central hole-spin}
Our microcavity-enhanced Raman control strategy faciliatates spin control with Rabi frequencies that can be precisely tuned over two orders of magnitude. For electron spins in both InGaAs 
\cite{Bodey2019} and GaAs QDs \cite{Nguyen2023}, Rabi frequencies close to the Larmor frequencies of the host nuclei enable single nuclear-spin flips mediated by the central spin (a Hartmann-Hahn resonance \cite{Hartmann1962}). Controlled activation of Hartmann-Hahn resonances has been used with great success to extend electron spin coherence times via nuclear bath cooling \cite{Gangloff2019, Jackson2022, Nguyen2023}; however, a hole spin has a more complex hyperfine interaction, and nuclear bath cooling has not yet been demonstrated.

\begin{figure*}[bth!]
    \centering
    \includegraphics[width = \linewidth]{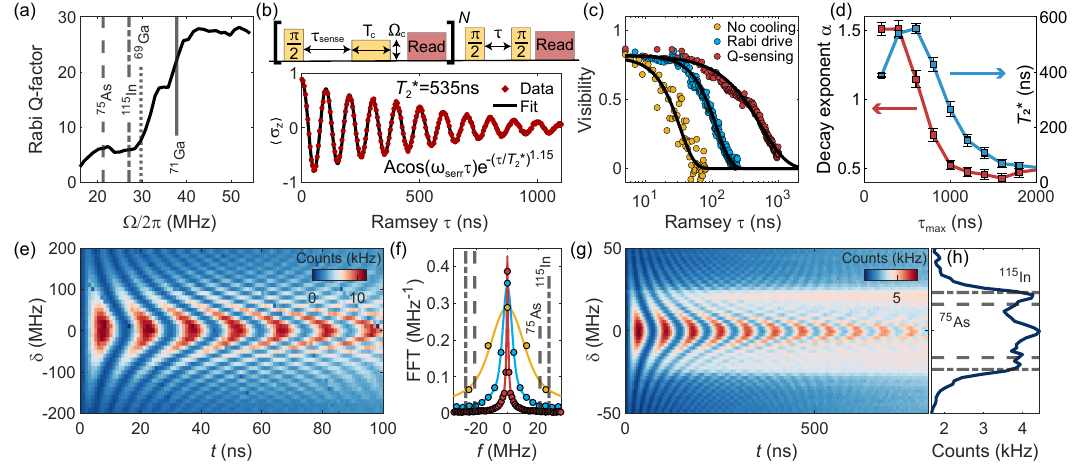}
    \caption{\textbf{Nuclear spin cooling mediated by a central hole spin.}
    \textbf{(a)} Hartmann-Hahn-like resonances observed in the $Q$-factor of the Rabi oscillations as $\Omega/2\pi$ is scanned through the nuclear Larmor frequencies.
    \textbf{(b)} Ramsey experiment to measure $T_2^*$ for quantum sensing-based nuclear cooling. Top: schematic of the cooling pulse sequence, which consists of $N$ repetitions of a cycle that senses and corrects nuclear spin fluctuations away from a target set point. A $\frac{\pi}{2}$ rotation brings the hole spin onto the equator of the Bloch sphere, followed by a sensing time $\tau_{\text{sense}}$, a Rabi drive $\Omega_c$ at the indium Larmor frequency, and a reset/readout pulse. Bottom: after cooling of the nuclei, the hole spin $T_2^*$ time is extended to 535\,ns.
    \textbf{(c)} Decay envelope of the hole spin coherence in a Ramsey experiment for no cooling, Rabi drive cooling and quantum sensing-based cooling. The fit function is $A\exp\left[-(t/T_2^*)^\alpha\right]$, where $A$ is the maximum visibility and $\alpha$ is a fit parameter. For no cooling and Rabi drive cooling, $\alpha=2$, whereas for sensing-based cooling $\alpha=1.15$.
    \textbf{(d)} Decay exponent $\alpha$ and $T_2^*$ as a function of $\tau_\text{max}$, the maximum sensing time in the cooling sequence. $T_2^*$ is maximised for $\tau_\text{max}=600$\,ns; interestingly $\alpha$ decreases below 1 for $\tau_\text{max}\gtrsim800$\,ns.
    \textbf{(e)} Rabi oscillations versus $\delta$ using Rabi drive cooling, showing a high-quality chevron pattern. 
    \textbf{(f)} Narrowing of the nuclear distribution with cooling (from an FFT of the Ramsey decay envelopes). Dashed lines show the expected positions to excite nuclear magnons; cooling brings the nuclear spin ensemble into the sideband-resolved regime.
    \textbf{(g)} Rabi chevron with sensing-based cooling and $\Omega/2\pi=13.6\text{MHz}<\omega_n/2\pi$; magnon sidebands are observed when the condition $\omega_n/2\pi = \sqrt{(\Omega/2\pi)^2+\delta^2}$ is satisfied. 
    \textbf{(h)} Averaged data from (g) over the range 315-500\,ns. Nuclear magnon sidebands are clearly observed.
    }
    \label{cooling}
\end{figure*}

We scan the hole spin Rabi frequency over the range of nuclear Larmor frequencies expected at 2.9\,T ($\omega_n/(2\pi)\approx20-50$~MHz). 
We find a non-trivial dependence of the oscillation quality factor ($Q = 2\cdot T_2^{\text{Rabi}}\cdot f_{\text{Rabi}}$) on $\Omega/(2 \pi)$, Fig.~\ref{cooling}(a). In particular, when the hole spin is driven close to the nuclear Larmor frequencies, the $Q$-factor drops. Similar results for electron spins in InGaAs and GaAs QDs have been attributed to Hartmann-Hahn resonances \cite{Bodey2019, Nguyen2023}; our results constitute a Hartmann-Hahn-like resonance on a QD hole spin.

For an electron spin, driving spin rotations at a Hartmann-Hahn resonance can be used to extend the electron-spin $T_{2}^{*}$ coherence time \cite{Gangloff2019,Jackson2022,Nguyen2023}. In fact, $T_{2}^{*}$ acts as a thermometer for the nuclear spins: a prolonged $T_{2}^{*}$ corresponds to a cooling of the nuclear spins, equivalently a narrowing of the Overhauser field distribution \cite{Gangloff2019}. We attempt the same scheme on a hole spin. We find that adding a Rabi drive prior to the Ramsey sequence increases the hole spin $T_{2}^{*}$ from $T_{2,\text{bare}}^{*}=28$~ns to $T_{2}^{*}=120$~ns. The Rabi drive of the hole spin results in nuclear spin cooling. Conversely, success here proves that $T_{2,\text{bare}}^{*}$ is limited by nuclear spin noise.

We attempt to cool the nuclei further using a quantum-sensing based scheme \cite{Jackson2022}; the pulse schematic in Figure~\ref{cooling}(b) shows the three-step cooling protocol. First, the hole spin is initialised and prepared in a superposition state with a $\frac{\pi}{2}$-pulse. The hole then senses the Overhauser field during a sensing period of $\tau_{\text{sense}}$. Second, a hole-nuclei flip-flop process is driven via a Hartmann-Hahn resonance, a process which flips a nuclear spin with a direction dependent on the phase accumulated by the hole spin during the sensing period. In this way, Overhauser field fluctuations away from the target value are corrected \cite{Jackson2022}. Third, an optical projective readout measurement removes entropy from the nuclei. This cycle is repeated $N$ times with increasing values of the sensing time $\tau_{\text{sense}}$. In this way, the feedback function is narrowed with each cycle and the sensitivity is increased. 
We find that this scheme is very effective in prolonging the $T_{2}^{*}$ of the hole spin. The optimal parameters in our case are $N=35$ with $\tau_{\text{sense}}$ linearly increasing from $\tau_{\text{min}} = 10$~ns to $\tau_{\text{max}}= 600$~ns, a Hartmann-Hahn drive with pulse length $T_c = 60$~ns at Rabi frequency 26~MHz and a readout pulse of 90~ns duration. Repeating the cooling cycle before every point in the Ramsey sequence shows a dramatic increase in coherence time from $T_{2, \text{bare}}^{*}=28$~ns to $T_{2, \text{cooled}}^{*}=535$~ns, Fig. \ref{cooling}\,(b). Probing the timescale over which the nuclear ensemble remains cold (by inserting a delay between the cooling sequence and the Ramsey experiment) results in no measurable decrease in the $T_2^*$ time up to the maximum probed delay of 1\,ms.

The Ramsey data in Fig.~\ref{cooling}\,(b) has a 10\,MHz modulation artificially added (via a $\tau$-dependent phase on the second $\pi$/2 pulse) to improve the fit quality; Fig. \ref{cooling}\,(c) shows the Ramsey decay envelope for both Rabi drive cooling and sensing-based cooling. The fit function is proportional to $\exp\left[-(\tau/T_2^*)^\alpha\right]$, where $\alpha$ is a free fit parameter. We observe an interesting dependence of $\alpha$ on $\tau_\text{max}$, the maximum sensing time in the cooling sequence. Figure \ref{cooling}\,(d) shows $\alpha$ (left axis) and $T_2^*$ (right axis, defined as the $1/e$ decay time) as a function of $\tau_\text{max}$. For $\tau_\text{max}>800$\,ns, we find that $\alpha$ decreases below 1, reaching a value of 0.5 for $\tau_\text{max}\gtrsim1000$\,ns. We note that the $\exp\left[-(\tau/T_2^*)^\alpha\right]$ fit function worsens for short $\tau\lesssim50$\,ns when $\tau_\text{max}\gtrsim800$\,ns. These observations point to changes in the spectral distribution of the spin noise after cooling.

We repeat a similar Rabi chevron experiment to Fig. \ref{concepts}\,(f), now with a cooling cycle before each data point. Figure~\ref{cooling}\,(e) shows the data using Rabi drive cooling; we observe a high-quality Rabi chevron, free from the background modulation observed in Fig.~\ref{concepts}\,(f). Despite the complexity of the semiconductor environment, the data closely matches the textbook result for a two-level system.

Following the analysis of Ref. \cite{Gangloff2019}, the effective nuclear temperature we achieve with the quantum-sensing based cooling scheme is $\sim180\,\mu$K. Such a low nuclear temperature brings us into the regime where collective nuclear excitations (i.e. magnons) can be observed \cite{Gangloff2019}. Figure~\ref{cooling}\,(f) shows the width of the Overhauser distribution for the thermal and cooled nuclear ensembles. We perform another Rabi chevron experiment, now with quantum-sensing based cooling before each data point and a low Rabi frequency ($\Omega<\omega_n$), Fig.~\ref{cooling}\,(g). In addition to the high-quality chevron, nuclear sidebands are observed when the Hartmann-Hahn resonance condition $\omega_n/2\pi = \sqrt{(\Omega/2\pi)^2+\delta^2}$ is satisfied. Figure~\ref{cooling}\,(h) shows the expected position of the $^{115}\text{In}$ and $^{75}\text{As}$ sidebands, consistent with our measured data.

\section{Hole spin discussion}
The nuclear-spin cooling protocol is clearly successful, prolonging $T_{2}^{*}$ from 28~ns to 535~ns. First, this proves that $T_{2}^{*}$ for uncooled nuclear spins is determined by nuclear-spin noise. If for instance $T_{2,\text{bare}}^{*}=28$~ns were limited by a different mechanism then the nuclear-spin cooling protocol would not increase $T_{2}^{*}$. Second, the success of the cooling protocol points to the presence of a non-colinear term in the hole-spin hyperfine interaction. This term allows a nuclear spin to be flipped with little energy cost. 

In general, $T_{2,\text{bare}}^{*}$ is determined by both nuclear-spin noise and charge noise (the latter via the dependence of the hole g-factor on electric field \cite{Prechtel2015}). The two processes have a different dependence on applied magnetic field, $B_o$. For large in-plane magnetic fields, spin noise results in a scaling of $T_{2}^{*} \propto B_o$. (This result is derived in the pure heavy-hole limit \cite{Fischer2008}.) The g-factor depends linearly on electric field \cite{Prechtel2015} such that charge noise leads to a scaling of $T_{2}^{*} \propto 1/B_o$. There is a spread of $T_{2}^{*}$ values in the literature but without a $B_o$-dependence or a characterisation of the charge noise, the mechanism limiting $T_{2}^{*}$ in most of these experiments is hard to pin down. However, Huthmacher {\em et al.} observed $T_{2}^{*} \propto B_o$ for $B_o$ up to about 3~T showing that spin noise is dominant at these fields \cite{Huthmacher2018}. At our Zeeman frequency (5.8~GHz), these authors find $T_{2}^{*} \simeq 70$~ns. This is higher than our own value, $T_{2}^{*}=28$~ns. This may be related to morphological differences, a different In composition for instance, and we note that the emission wavelengths are quite different (923~nm here, 970~nm in Ref.~\cite{Huthmacher2018}).

The leading term in the hole hyperfine Hamiltonian has an Ising form, $S_z I_z$, and arises from the dominant heavy-hole contribution to the hole state \cite{Fischer2008}. The full Hamiltonian is not known with any precision. However, theory for an InGaAs quantum dot predicts the existence of a non-colinear term \cite{Ribeiro2015}. Our experiment confirms its presence via the observation of hole-spin-driven nuclear-spin cooling and the observation of magnons.

We now address the Hahn echo. Our $T_{2}^{\text{HE}}$ is considerably larger than previous values. This points to the importance of charge noise: our $T_{2}^{\text{HE}}$ is large because charge noise is small. There is some evidence that charge noise limits our $T_{2}^{\text{HE}}$. The Hahn echo decay, $\exp[-(\tau/T_{2}^{\text{HE}})^{\alpha})]$, fits well with $\alpha \approx 2$ but poorly with $\alpha=3$: this is consistent with an approximate $1/f$-dependence (and not a $1/f^2$-dependence \cite{Medford2012}) of the noise spectrum $S(f)$. A $1/f^{\beta}$dependence with $\beta \approx 0.5-1$ is typical of charge noise \cite{Kuhlmann2013}. Furthermore, the scaling of $T_2$ with the number ($N$) of CPMG pulses, $T_2 = T_{2}^{\text{HE}}\cdot N^{\gamma}$, gives $\gamma=0.31$, corresponding to $S(f) \propto 1/f^{\gamma/1-\gamma} \approx 1/f^{0.45}$, again characteristic of charge noise.

CPMG with just 8 pulses brings $T_2$ to the limit determined by spin relaxation, i.e., $T_2=2 T_1$ with $T_1=20$~$\mu$s. The mechanisms limiting $T_1$ are not known. One possible contribution comes from phonon scattering, turned on via a spin-orbit interaction \cite{Trif2009}. In the limit of small $Z_h$, the hole-spin $T_1$ was measured to be of order 100 $\mu$s at 4.2~K \cite{Gerardot2008,Finley2007}. This is larger but nevertheless comparable (given the strong dependence of $T_1$ on the morphology) to the hole spin $T_1$ measured here. Another possible contribution to our hole-spin $T_1$ is escape of the metastable hole from the QD \cite{Lochner2021, Mannel2023}.

A previous spectroscopy experiment (coherent population trapping, CPT) determined a very small uncertainty in the hole Zeeman frequency, equivalently $T_{2}^{*}=460$~ns for Zeeman splittings of 1.5--2.65~GHz \cite{Prechtel2016}. The experiment was performed on a low-charge-noise device -- without this, such a large $T_{2}^{*}$ would be impossible. $T_{2}^{*}=460$~ns is much larger than $T_{2, \text{bare}}^{*}$ reported here yet similar to the value recorded after full nuclear-spin cooling. The implication is that the nuclear spins were (unknowingly) cooled in the CPT experiment. This idea is given some support from the fact that CPT is known to result in nuclear-spin cooling for well-chosen parameters \cite{Ethier-Majcher2017}.

The extremely slow heating rate of the cooled nuclear spins points to a suppressed energy exchange between the nuclear spins in the QD and those in the environment. This mimics the extremely long decay of nuclear spin polarisation in this system \cite{Maletinsky2007,Wust2016}. It is likely therefore that the slow energy exchange has the same explanation: strain makes the nuclear spins in the QD energetically different from those outside the QD (via the quadrupolar interaction \cite{Stockill2016}), suppressing spin diffusion.

\section{Conclusions}
We have demonstrated coherent control of a hole spin in an optical microcavity. We exploit cavity-enhancement of the Raman process to achieve very high Rabi frequencies, with couplings larger than 2~GHz. Hahn echo gives a $T_{2}^{\text{HE}}$ of 20~$\mu$s, a record-high value for a spin in an InGaAs QD. For the first time, we implement nuclear-spin cooling schemes using the hole as central spin, thereby extending the hole spin $T_{2}^{*}$ to 535~ns. The coherence time $T_{2}^{*}$ and the spin rotation time are both much larger than the recombination time (50~ps in the microcavity) allowing many photons to be created before spin coherence is lost. The combination of high spin coherence times, fast spin manipulation, high end-to-end photon efficiency \cite{Tomm2021}, fast initialisation (3~ns with fidelity $F \geq 96.7$\%), and fast single-shot spin readout \cite{Antoniadis2023} highlights the potential of this platform. The Purcell-enhanced branching ratio facilitates the creation of a spin-photon interface, also a source of entangled cluster states, in the time-bin basis \cite{Tiurev2022}. 

\section{Acknowledgments}
We acknowledge financial support from Horizon 2020 FET-Open Project QLUSTER and Swiss National Science Foundation project 200020\_204069. G.N.N. and A.J.\ acknowledge support from the European Union's Horizon 2020 Research and Innovation Programme under Marie Sk\l{}odowska-Curie grant Agreement No. 861097 (QUDOT-TECH) and No.\ 840453 (HiFig) respectively. S.R.V., R.S., A.L.\ and A.D.W.\ gratefully acknowledge support from DFH/UFA CDFA05-06, DFG via ML4Q EXC 2004/1 - 390534769, and the BMBF project QR.X 16KISQ009.\\

\section{Methods}
Our system consists of an InGaAs/GaAs heterostructure containing the QDs and a highly reflective distributed Bragg reflector which forms the bottom mirror of the microcavity. The QDs lie in the intrinsic region of an \textit{n-i-p} diode structure with a tunnel coupling to the $n$-doped layer.  Each QD operates in the Coulomb blockade regime. Co-tunneling results in a short electron-spin relaxation time $T_1$ in this particular device \cite{Antoniadis2023}. Here, we operate instead with a single hole. By applying an appropriate voltage and in the presence of many holes, Coulomb blockade ensures that just one hole is trapped on the QD. The top mirror is a curved crater in a silica substrate (fabricated by laser ablation \cite{Hunger2012}) coated with a Bragg mirror, 8 layer pairs of Ta$_2$O$_5$/SiO$_2$. The top mirror is fixed in position and the heterostructure can be moved relative to it in situ via XYZ nanopositioners. This allows the microcavity mode to be optimally coupled to the QD (by adjusting X and Y), and the microcavity resonance to be coupled to a transition of choice (by adjusting Z). The reflectivity of the top mirror is chosen to maximise the conversion probability of a QD-exciton to photon exiting the microcavity \cite{Tomm2021}. The microcavity is one-sided: laser excitation and single-photon extraction both proceed via the top mirror as shown schematically in Fig.~\ref{concepts}(a). A cross-polarised dark-field microscope separates laser excitation and QD photons \cite{Kuhlmann2013}. The microcavity mode is split into two orthogonal, linearly-polarised modes; the splitting is 50~GHz. As a deterministic single-photon source, the end-to-end efficiency (the probability of creating a single photon at the output of the collection fibre following a trigger, an excitation pulse) of the same system was reported as 57\% \cite{Tomm2021}. Here, the end-to-end efficiency is slightly less (estimated 35\%) as there are some losses at an additional beam-splitter which couples the Raman laser to the on-axis optical mode (see Supplementary Note 1). Between source and detector we insert a transmission filter (transmission 50\%) to remove the Raman laser. The microcavity setup is placed into a helium bath cryostat at 4.2~K. 

\clearpage
\onecolumngrid
\renewcommand{\thefigure}{S\arabic{figure}}
\def\thesection{Supplementary Note \arabic{section}}
\def\thesubsection{\arabic{section} \arabic{subsection}}
\setcounter{section}{0}
\setcounter{figure}{0}

\renewcommand{\thetable}{S\arabic{table}}  

\renewcommand{\figurename}{\MakeUppercase{Supplementary Fig.}}
\renewcommand{\thesubsection}{{Supplementary Note} \arabic{section}.\arabic{subsection}}

\section*{Supplementary Information: \\Fast optical control of a coherent hole spin in a microcavity}
	
\section{Optical setup and microwave-modulated spin control}
\begin{figure}[b]
    \centering
    \includegraphics[width = 0.9\linewidth]{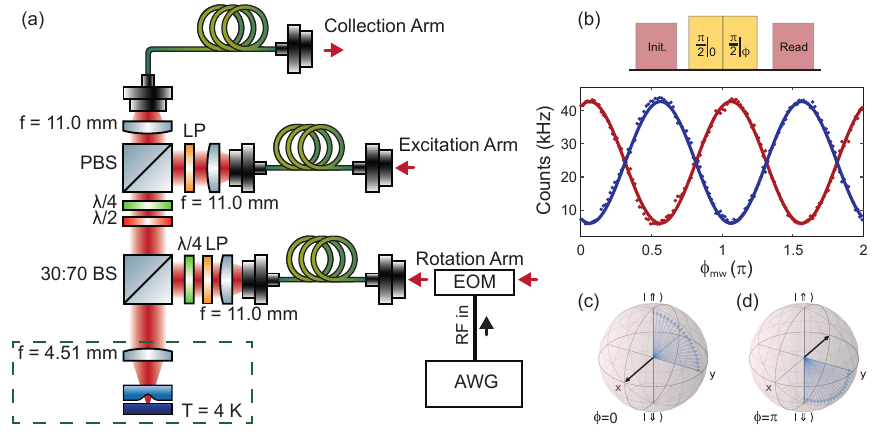}
    \caption{\textbf{Experimental setup.}
    \textbf{(a)} Schematic of the optical setup used for QD spin control. The spin control laser is coupled to the optical axis of the microscope head using a 30:70 (R:T) beamsplitter; this is the only modification required to integrate coherent spin control with our efficient microcavity single-photon source \cite{Tomm2021}.
    \textbf{(b)} Flexible control of the rotation axis for the QD spin: two $\pi/2$ pulses are applied with zero delay between them. The microwave phase ($\phi_\text{mw}$) of the second pulse is swept from 0 to 2$\pi$ (red data points, blue data points have a second pulse phase at the qubit of $\phi+\pi$). The signal oscillates as the rotation axis of the second pulse is swept around the equator of the Bloch sphere.
    \textbf{(c)} Bloch sphere schematic of the second $\pi/2$ pulse for $\phi=0$: here, the two $\pi/2$ pulses both induce a rotation around the $x$-axis, and combine to create a $\pi$-pulse.
    \textbf{(d)} Bloch sphere schematic for $\phi=\pi$: here the second $\pi/2$ pulse brings the spin back to the initial state.
    }
    \label{fig:phase}
\end{figure}

\noindent
Figure \ref{fig:phase}\,(a) shows a schematic of the microscope head used to excite the cavity-quantum dot system and to collect the emitted photons. Resonant spin initialisation and readout pulses are applied via the excitation arm, and photons emitted by the quantum dot are collected using a cross-polarised dark-field microscope to filter the excitation laser \cite{Kuhlmann2013}. A third arm (the rotation arm) is used to couple the circularly polarised, red-detuned two-colour Raman laser used for coherent spin control into the beam column via a 30:70 (R:T) beamsplitter.

Coherent spin control is achieved using a two-colour Raman scheme \cite{Bodey2019}. We implement this concept by modulating the output of a single-frequency CW laser (red detuned by $\Delta$ from the readout laser) with an amplitude electro-optic modulator (EOM) driven by a fast AWG (Tektronix AWG70000B, which also controls the AOM used to generate initialisation/readout pulses). 
The AWG provides a pulsed sinusoidal signal with frequency $f_\text{mw}$, which generates two sidebands in the optical signal at $f_\text{Raman}\pm f_\text{mw}$ (with phase $\pm\phi_\text{mw}$). The DC bias of the EOM is continuously stabilised to suppress the carrier signal at $f_{\rm Raman}$, such that in the ideal case the output from the EOM consists only of the two sidebands. When $2f_{\rm mw}=Z_h$ (where $Z_h$ is the hole spin Zeeman splitting) the two sidebands drive a coherent simulated Raman transition between the two spin states \cite{Bodey2019}. 

The EOM imprints the microwave signal generated by the AWG onto the optical fields: by adjusting the phase of the microwave signal from the AWG ($\phi_\text{mw}$), we can achieve arbitrary single-qubit rotations of the hole spin. 
Figure \ref{fig:phase}\,(b) demonstrates control of the hole spin rotation axis on the Bloch sphere. Here, initialisation into the $\ket{\Downarrow}$ state is followed by a $\pi/2$ rotation around the $x$-axis to create a superposition state. We then apply a second $\pi/2$ rotation with a phase $\phi_\text{mw}$ relative to the first pulse, followed by a readout pulse. 
The red data points in Fig.\ \ref{fig:phase}\,(b) show the readout signal as a function of $\phi_\text{mw}$, which oscillates as the rotation angle of the second pulse moves around the equator of the Bloch sphere. We note that the pulse phase experienced by the qubit is $\phi=2\phi_\text{mw}$ on account of the two optical sidebands with phase $\pm\phi_\text{mw}$ (explaining why the oscillations in Fig.\ \ref{fig:phase}(b) are $\pi$-periodic rather than $2\pi$-periodic). In the main text all pulse phases are expressed as that phase experienced by the qubit (so e.g. the $\pi_\text{y}$ pulses in a CPMG sequence are offset in phase by $\phi=\pi/2$  relative to the first pulse, corresponding to $\phi_{mw}=\pi/4$). We can thus achieve multi-axis control of our hole spin qubit by simply adjusting the AWG output waveform, allowing more complex pulse sequences to be implemented easily. The blue data points have a second pulse phase (as experienced by the qubit) of $\phi+\pi$, which inverts the $z$-projection of the spin relative to the red data points.

%%%%%%%%%%%%%%%%%%%%%%%%%%%

\section{Detuning dependence of the optical enhancement for a one-sided cavity}

\noindent In our system, the rotation laser used to control coherently  the QD spin is in-coupled via the top mirror of the microcavity. This approach has the advantage of simplicity: the rotation laser is aligned to exactly the same beam path as the resonant readout laser, and no modifications are required to the cavity structure (which is identical to that previously used for high-efficiency single-photon generation \cite{Tomm2021}). One concern with this approach is that the cavity should be resonant with the readout transition for optimal photon collection, whereas the spin rotation laser must be sufficiently red detuned to avoid significant population of the exciton states during the rotation pulses (which leads to incoherent photon scattering and randomisation of the spin state). Thus the spin rotation laser must be detuned from the cavity resonance, and, depending on the detuning, may be filtered by the cavity.

Here, we derive the expected field enhancement of a one-sided cavity as a function of detuning, and show that our cavity parameters (which are optimal for a single-photon source) are also highly compatible with all-optical spin control. In fact, the cavity plays an advantageous role in \emph{enhancing} the optical field of the rotation laser, enabling ultra-fast Rabi rotations. We note that because the spin control scheme relies on a two-photon process, the spin Rabi frequency is proportional to the optical intensity rather than the electric field: the intensity enhancement is thus ultimately the quantity we are interested in.

\begin{figure}[b]
    \centering
    \includegraphics[width=\linewidth]{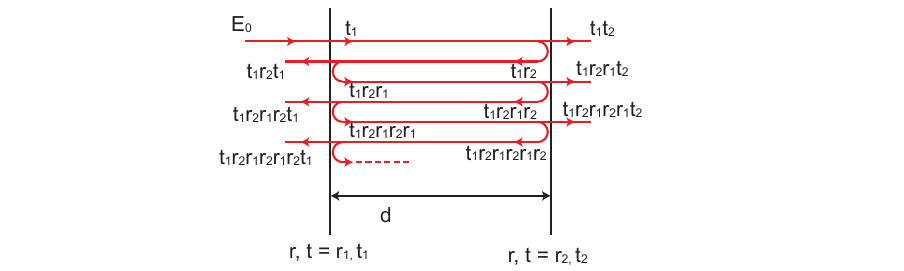}
    \caption{\textbf{Cavity field schematic.} 
    Illustration of the transmission $t_i$ and reflection $r_i$ coefficients of an incoming electric field $E_0$ incident on a cavity consisting of two mirrors separated by distance $d$. The sketch shows the field components for the first few reflections.}
    \label{fig:deriv}
\end{figure}

We begin with a simple sketch of a light field with electric field amplitude $E_0$ incident on a cavity as shown in Fig.\ \ref{fig:deriv}. The derivation is a modification of Ch.\,3 of Ref.\ \cite{Nagourney2010} for a one-sided cavity.
The ratio of the electric field at the centre of the cavity to the input field can be calculated by adding up all left and right propagating components inside the cavity:
\begin{align}
    \frac{E_c}{E_0} &= t_1 e^{-i\delta/4} + t_1\cdot r_2 e^{-3i\delta/4} + t_1\cdot r_2\cdot r_1 e^{-5i\delta/4} + t_1\cdot r_2\cdot r_1 \cdot r_2 e^{-7i\delta/4} + ... \nonumber
    \\ &= t_1 e^{-i\delta/4}\left[\frac{1+ r_2 e^{-i\delta/2}}{1- r_1r_2 e^{-i\delta}}\right],
    \label{eq:deriv_Ec}
\end{align}
where $\delta$ is the phase change acquired on a round-trip and $r_i/t_i$ are the reflection/transmission coefficients of the two mirrors. 
From Eq.\ \ref{eq:deriv_Ec} we can calculate the ratio of the full field intensity at the centre of the cavity and the intensity of the input field:
\begin{equation}
    \frac{I_c}{I_0} = \left|\frac{E_c}{E_0}\right|^2 = \frac{(1-r_1^2)(r_2^2 +1 + 2r_2\cos(\delta/2))}{(1-r_1r_2)^2 + 4r_1r_2\sin^2(\delta/2)} \approx \frac{t_1^2(r_2+1)^2}{(1-r_1r_2)^2 + r_1r_2(\delta-\delta_0)^2},
    \label{eq:deriv_Ic}
\end{equation}
where the approximation is valid close to the cavity resonance (occurring at $\delta=\delta_0$). Several parameters can be extracted from Eq.\ \ref{eq:deriv_Ic}. The resonances of the intensity in the cavity are found at $\sin(\delta_0/2) = 0$, hence $\delta_0 = j\cdot 2\pi$ ($j \in \mathbb{N}$) and the free spectral range (FSR) of the cavity is ${\rm FSR}_{\delta} = 2\pi$. Using the small angle approximation we find the half of the maximum intensity at $(1-r_1r_2)^2 = r_1r_2(\delta-\delta_0)^2$. Therefore, $(\delta-\delta_0) = (1-r_1r_2)/\sqrt{r_1r_2}$ leading to a full-width-at-half-maximum ${\rm FWHM}_{\delta} \approx 2(1-r_1r_2)$. 

Using these equations for ${\rm FWHM}_{\delta}$ and ${\rm FSR}_{\delta}$, we define the finesse of the cavity in terms of the reflection and transmission coefficients:
\begin{equation}
\mathcal{F} = \frac{\mathrm{FSR}_{\delta}}{\mathrm{FWHM}_{\delta}} = \frac{\pi}{1-r_1r_2}. \end{equation}

Some experimental parameters are connected with these expressions. According to Eq.\ \ref{eq:deriv_Ic}, the ${\rm FWHM}_{\delta} = \Delta \delta = 2(1-r_1r_2) = 2d\kappa/c$, with $c$ the speed of light and $\kappa/(2\pi)$ the cavity linewidth. Therefore, $\delta = 4 \pi d/\lambda = 4\pi d f/c$, where $f$ is the laser frequency. Further, ${\rm FSR}_f = \mathcal{F}\cdot\kappa/(2\pi) = c/(2 d)$. (${\rm FSR}_f$ is the FSR in the frequency domain.) 

In order to estimate the detuning dependence of the field enhancement in the cavity we plug these relations into Eq.\,\ref{eq:deriv_Ic} for a one-sided cavity, where where $r_2 = 1$, and $r_1 = r$:
\begin{equation}
    \frac{I_c}{I_0} = \frac{4t_1^2}{(1-r)^2 + r(\delta-\delta_0)^2} \approx \left(\frac{2\mathcal{F}t}{\pi}\right)^2 \frac{(\kappa/4\pi)^2}{(\kappa/4\pi)^2 + (f-f_0)^2} = X\frac{(\kappa/4\pi)^2}{(\kappa/4\pi)^2 + (f-f_0)^2},
    \label{eq:deriv_detuning_Ic}
\end{equation}
where $X$ is the intensity enhancement at the cavity resonance. On resonance of Eq.\ \ref{eq:deriv_Ic} ($\delta-\delta_0 = 0$):
\begin{equation}
    \frac{I_c}{I_0} = \frac{4(1- r^2)}{(1-r)^2} =   \frac{4(1+r)}{(1-r)} \approx 8 \mathcal{F}/\pi = X.
\end{equation}
The enhancement at resonance with the cavity, $X$, is thus given by $8 \mathcal{F}/\pi$. 
In our experiments, $\mathcal{F} = 500$, hence $X = 1,270$. 
In order to determine the detuning dependence of the enhancement in our experimental case, we take Eq.\ \ref{eq:deriv_detuning_Ic} and calculate the detuning at which the cavity enhancement is equal to one (i.e., the intra-cavity intensity is equal to the incident intensity).
We assume $\left|f - f_0\right| \gg \kappa/(2\pi)$. In that case
\begin{equation}
\frac{I_c}{I_0} \approx X \cdot \frac{(\kappa/4\pi)^2}{(f-f_0)^2},  
\end{equation}
i.e., $I_c = I_0$ for 
\begin{equation}
    f-f_0 = \frac{\kappa}{4\pi}\sqrt{\frac{8F}{\pi}} = \frac{\kappa}{2\pi}\sqrt{\frac{2F}{\pi}} \approx 18 \frac{\kappa}{2\pi} = 450~\mathrm{GHz}.
\end{equation}
Hence, for our cavity linewidth $\kappa/(2\pi) = 25$~GHz, this condition applies at a detuning of 450~GHz. In other words, for detunings up to 450~GHz, the rotation laser will be enhanced by the cavity -- only beyond 450~GHz will cavity suppression be present. 

In order to avoid the rotation laser populating the exciton states, the detuning should be much larger than the spin Rabi frequency. Detunings of 450~GHz are suitable for driving spin rotations -- experimentally, we find an optimum rotation fidelity in a broad range around 320~GHz detuning, well inside the cavity enhancement regime. 

\begin{figure}[t]
    \centering
    \includegraphics[width=\linewidth]{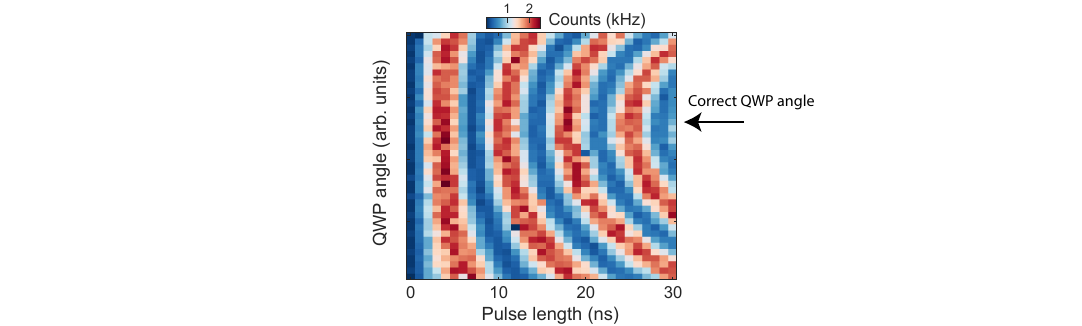}
    \caption{\textbf{Calibration of the rotation laser polarisation.} 
    We achieve circular polarisation at the QD position by driving Rabi oscillations of the hole spin as a function of the QWP angle in the rotation laser arm. For a fixed rotation laser power, the Rabi frequency is a direct measurement of the polarisation experienced by the QD. The optimal polarisation maximises the spin Rabi frequency, indicated by the black arrow. We note that depending on $\Delta$, the input polarisation should be slightly elliptical to produce circular polarisation at the QD position. Experimentally this is trivial to account for, simply by re-calibrating the QWP angle when changing $\Delta$.
    }
    \label{fig:qwp}
\end{figure}

We note that our cavity has two non-degenerate modes with orthogonal (H and V) polarisation and a mode splitting of 50~GHz \cite{Tomm2021}. This means that for a given rotation-laser detuning $\Delta$, the field enhancement has a slight polarisation dependence. However, this polarisation dependence presents no issue for spin control: the rotation laser-field experienced by the QD should have circular polarisation, which is achieved by sending in slightly elliptically polarised light to counteract the polarisation-dependent field enhancement. The optimal polarisation is calibrated using the fact that for ideal circular polarisation at the QD position, the spin Rabi frequency is maximised. An example calibration experiment is shown in Fig.\ \ref{fig:qwp}, here for a detuning $\Delta=280$~GHz. The correct quarter-wave-plate (QWP) angle for this detuning is indicated by the black arrow. On changing $\Delta$, the QWP angle should be re-calibrated. In this way circular polarisation at the QD position is achieved in the experiment.
 
%%%%%%%%%%%%%%%%%%%%%%%%%%
\section{Spin initialisation and readout}
\begin{figure}[b]
    \centering
    \includegraphics[width = 0.9\linewidth]{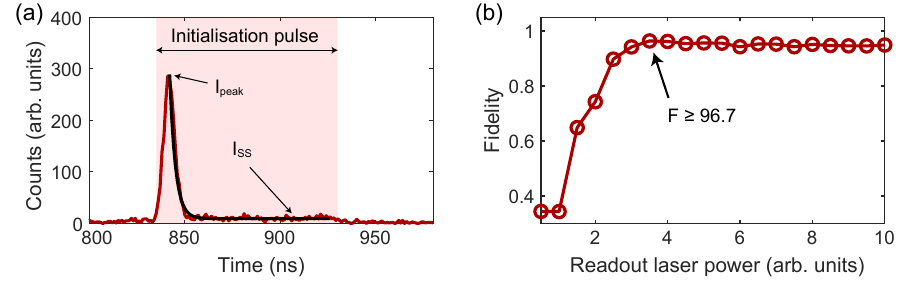}
    \caption{\textbf{Spin initialisation fidelity.}
    \textbf{(a)} Histogram of an initialisation pulse. A prior preparation sequence prepares $\ket{\Uparrow}$, the bright state (we assume with unit fidelity, giving us a lower bound on the initialisation fidelity extracted from this experiment). The initialisation pulse pumps the spin into state $\ket{\Downarrow}$. The initially high counts $I_{\rm peak}$ decay exponentially to a steady-state level $I_{\rm ss}$. A lower bound to the initialisation fidelity can be calculated from $I_{\rm peak}$ and $I_{\rm ss}$. An initialisation time about 3~ns is extracted from the exponential decay.
    \textbf{(b)} Initialisation fidelity $F$ for different laser powers. A lower bound of $F \geq 96.7$\% is extracted.}
    \label{fig:init}
\end{figure}

\noindent
All pulse sequences used in this work start with a pulse that initialises the spin state, typically spin down $\ket{\Downarrow}$. This is achieved by resonantly driving the $\ket{\Uparrow} \leftrightarrow \ket{\Uparrow\Downarrow, \uparrow}$ transition, which can decay to either $\ket{\Uparrow}$ or $\ket{\Downarrow}$. In bulk structures the two decay paths are equally likely; our cavity can selectively enhance one of the transitions to create an asymmetric decay rate (of up to 1:10). Spin initialisation via optical pumping is achieved with high fidelity in our system: a lower bound for this initialisation fidelity can be estimated from the transient spin-pumping signal measured during the initialisation pulse, as shown in  Fig.\ \ref{fig:init}(a). To perform this experiment, some spin population must first be transferred to the $\ket{\Uparrow}$ state (which gives fluorescence when driven by the initialisation laser, hence is termed the ``bright'' state). The initialisation laser by default initialises $\ket{\Downarrow}$ (the ``dark'' spin state); population can be transferred to the bright state with either a subsequent $\pi$-pulse or a second laser on resonance with the dark state. 

During the initialisation pulse, a transient signal with maximum signal $I_{\rm peak}$ is observed, exponentially decaying to a background level $I_{\rm ss}$. The values of $I_{ss}$ and $I_{\rm peak}$ can be extracted from an exponential fit. The initialisation fidelity is then given by \cite{Appel2021}:
\begin{equation}
    F = 1 - \rho_{11}(0) \cdot \frac{I_{\rm ss}}{I_{\rm peak}} + \rho_{\rm 11}(0) \cdot \Theta \cdot \frac{\gamma_x}{\gamma_0} \frac{I_{\rm ss}}{I_{\rm peak}}
    \label{eq:init}
\end{equation}
where $\rho_{11}(0)$ is the initial population of the bright spin state before the initialisation pulse is applied, $\gamma_{x,y}$ the emission rates from the excited trion state into $\ket{\Uparrow},\ket{\Downarrow}$ (labelled as states 1, 2 respectively), $\gamma_0 = \gamma_x + \gamma_y$ and $\Theta = \frac{\rho_{22}}{\rho_{11} + \rho_{22}}$. We can extract a lower bound on the initialisation fidelity by assuming $\rho_{11}=1$ (corresponding to perfect initial preparation of the bright state with no error) and neglecting the last term in Eq.\ \ref{eq:init}, using the values for $I_{\rm peak}$ and $I_{\rm ss}$ extracted from the data in Fig.\ \ref{fig:init}(a). 

The initialisation fidelity depends on the power of the initialisation laser for a given initialisation pulse length. At low laser powers, complete optical pumping is not achieved during the initialisation pulse. After reaching a maximum for increasing power, the fidelity stays more or less constant for a range of powers. However, for even higher powers, the fidelity reduces again, likely due to optical re-pumping. The lower bound of the fidelity for different readout powers is calculated via Eq.\ \ref{eq:init} and is shown in Fig.\ \ref{fig:init}b. We find $F \geq$ 96.7\% at the optimum power, with a fast initialisation time of $\sim 3$~ns.

Spin readout is performed using exactly the same pulse as spin initialisation, as the transient spin-pumping signal shown in Fig. \ref{fig:init} is proportional to the $\ket{\Uparrow}$ state population. Integrating the signal over a window covering the transient pulse signal provides our readout signal.

%%%%%%%%%%%%%%%%%%%%%%%%%%%

\section{Hole spin T$_1$ time}

\begin{figure}[b]
    \centering
    \includegraphics[width = \linewidth]{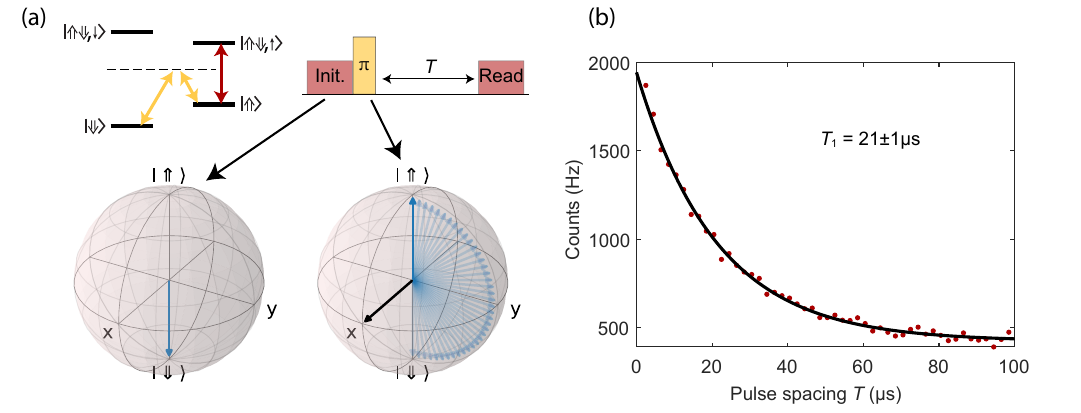}
    \caption{\textbf{Spin relaxation rate.}
    \textbf{(a)} Pulse sequence used to measure the hole spin $T_1$ time. The spin is prepared in state $\ket{\Uparrow}$ by initialising state $\ket{\Downarrow}$ using optical pumping, followed by a $\pi$-pulse. After a time $T$ the readout laser probes the remaining population in state $\ket{\Uparrow}$, thus measuring the spin relaxation rate.
    \textbf{(b)} Experimental data obtained using the pulse sequence in (a). Fitting to an exponential decay returns $T_1=21\pm1\mu$s.
    }
    \label{fig:t1}
\end{figure}

\noindent
We measure the relaxation rate ($T_1$) of the hole spin using the pulse sequence shown in Fig.\ \ref{fig:t1}(a). The spin is first initialised in state $\ket{\Downarrow}$ by resonantly driving the $\ket{\Uparrow}\leftrightarrow\ket{\Uparrow\Downarrow,\uparrow}$ transition, which prepares state $\ket{\Downarrow}$ by optical pumping as described in the previous section. We then apply a $\pi$-pulse to the hole spin to prepare the state $\ket{\Uparrow}$. After a wait time $T$ we perform a readout pulse that measures the population remaining in state $\ket{\Uparrow}$ (again by driving the $\ket{\Uparrow}\leftrightarrow\ket{\Uparrow\Downarrow,\uparrow}$ transition, which also prepares $\ket{\Downarrow}$ for the next measurement cycle). By performing this sequence as a function of $T$, we can determine the rate at which the state $\ket{\Uparrow}$ relaxes to thermal equilibrium. Figure \ref{fig:t1}(b) shows the resulting data, which on fitting to an exponential decay gives $T_1=21\pm1\,\mu$s.

%%%%%%%%%%%%%%%%%%%%%%%%%%%
\section{Laser-induced spin flips}
\begin{figure}[h]
    \centering
    \includegraphics[width = 0.9\linewidth]{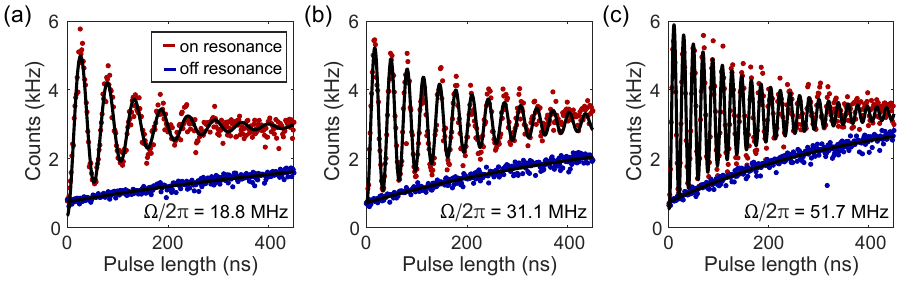}
    \caption{\textbf{Incoherent rotation laser-induced spin flips.} 
    Readout signal as a function of Rabi pulse duration $T$ on-(red) and off-(blue) resonant with respect to the hole Zeeman transition at a rotation laser detuning of 320~GHz at Rabi frequencies of  
    \textbf{(a)} $\Omega/(2 \pi)$ = 18.5~MHz,
    \textbf{(b)} $\Omega/(2 \pi)$ = 31.1~MHz, and
    \textbf{(c)} $\Omega/(2 \pi)$ = 51.7~MHz. For the highest Rabi frequency, the off-resonant curve approaches the envelope of the on-resonant curve, meaning that the incoherent spin-flip process becomes a limiting factor. 
    }
    \label{fig:spinflip}
\end{figure}

\noindent
At high rotation laser power, the Rabi oscillation $Q$-factor is limited by laser-induced incoherent spin flips. This effect can be quantified by comparing the readout signal for a Rabi experiment with the microwave drive on- and off-resonance with respect to the hole Zeeman splitting. Figures \ref{fig:spinflip}(a), (b) and (c) show the on- and off-resonant counts for $\Omega/(2\pi) =$ 18.5~MHz, 31.1~MHz and 51.7~MHz Rabi frequency, respectively, at a rotation laser detuning of 320~GHz. For large rotation pulse durations, the readout signal increases even in the off-resonant curves, a consequence of incoherent spin-flips populating the bright state. This process limits the $Q$-factor of the Rabi oscillations for high Rabi frequencies. The Raman-laser-induced spin-flip rate is linearly proportional to the Raman laser power, and is extracted from an exponential fit to the off-resonant signal. As the spin Rabi frequency is also linearly proportional to the Raman laser power, a ratio of the off-resonant spin flip rate scaled to the Rabi frequency can be defined. We find normalised rates in the range of $0.5-1.2 \times 10^{-4}$~ns$^{-1}$/MHz, with a weak dependence on $\Delta$. Previous work has observed significantly higher laser-induced spin flip rates for a hole spin ($\sim2.5-3.6\times10^{-4}\,\rm{ns^{-1}/MHz}$ \cite{Appel2022}) compared to an electron spin ($\sim0.2-0.4\times10^{-4}\,\rm{ns^{-1}/MHz}$ \cite{Nguyen2023, Bodey2019}). Interestingly our results for a hole spin show significantly lower laser-induced spin-flip rates than previous hole spin results, comparable to that for an electron spin. The mechanism leading to these incoherent spin-flips is not fully understood. 
 
\bibliography{references}

\end{document}